\documentclass[english,journal=jctcce,manuscript=letter,layout=twocolumn]{achemso}
\usepackage[T1]{fontenc}
\usepackage[utf8]{inputenc}
\usepackage{babel}
\usepackage{prettyref}
\usepackage{amsbsy}
\usepackage{amstext}
\usepackage{graphicx}
\usepackage[numbers]{natbib}
\usepackage[unicode=true,pdfusetitle,
 bookmarks=true,bookmarksnumbered=false,bookmarksopen=false,
 breaklinks=false,pdfborder={0 0 1},backref=false,colorlinks=false]
 {hyperref}

\makeatletter

\title{On the accuracy of a recent regularized nuclear potential}

\author{Susi Lehtola}

\email{susi.lehtola@alumni.helsinki.fi}

\alsoaffiliation{Department of Chemistry, University of Helsinki, P.O. Box 55, FI-00014
University of Helsinki, Finland}

\providecommand{\tabularnewline}{\\}

\usepackage[version=4]{mhchem}
\SectionNumbersOn
\usepackage{doi}

\usepackage{cleveref}
\AtBeginDocument{%
\let\ref\cref
}

\makeatother

\begin{document}
\begin{abstract}
F. Gygi recently suggested an analytic, norm-conserving, regularized
nuclear potential to enable all-electron plane-wave calculations {[}J.
Chem. Theory Comput. 2023, 19, 1300--1309{]}. This potential $V(r)$
is determined by inverting the Schrödinger equation for the wave function
ansatz $\phi(\boldsymbol{r})=\exp[-h(\boldsymbol{r})]/\sqrt{\pi}$
with $h(\boldsymbol{r})=r\text{erf}(ar)+b\exp(-a^{2}r^{2})$, where
$a$ and $b$ are parameters. Gygi fixes $b$ by demanding $\phi$
to be normalized, the value $b(a)$ depending on the strength of the
regularization controlled by $a$. We begin this work by re-examining
the determination of $b(a)$ and find that the original 10-decimal
tabulations of Gygi are only correct to 5 decimals, leading to normalization
errors in the order of $10^{-10}$. In contrast, we show that a simple
100-point radial quadrature scheme not only ensures at least 10 correct
decimals of $b$, but also leads to machine-precision level satisfaction
of the normalization condition.

Moreover, we extend Gygi's plane-wave study by examining the accuracy
of $V(r)$ with high-precision finite element calculations with Hartree--Fock
and LDA, GGA, and meta-GGA functionals on first- to fifth-period atoms.
We find that although the convergence of the total energy appears
slow in the regularization parameter $a$, orbital energies and shapes
are indeed reproduced accurately by the regularized potential even
with relatively small values of $a$, as compared to results obtained
with a point nucleus. The accuracy of the potential is furthermore
studied with $s$-$d$ excitation energies of Sc--Cu as well as ionization
potentials of He--Kr, which are found to converge to sub-meV precision
with $a=4$. The findings of this work are in full support of Gygi's
contribution, indicating that all-electron plane-wave calculations
can be accurately performed with the regularized nuclear potential.
\end{abstract}
\newcommand*\ie{{\em i.e.}}
\newcommand*\eg{{\em e.g.}}
\newcommand*\etal{{\em et al.}}
\newcommand*\citeref[1]{ref. \citenum{#1}}
\newcommand*\citerefs[1]{refs. \citenum{#1}} 

\newcommand*\Erkale{{\sc Erkale}}
\newcommand*\HelFEM{{\sc HelFEM}}
\newcommand*\Bagel{{\sc Bagel}}
\newcommand*\FHIaims{{\sc FHI-aims}}
\newcommand*\LibXC{{\sc LibXC}}
\newcommand*\Orca{{\sc Orca}}
\newcommand*\PySCF{{\sc PySCF}}
\newcommand*\PsiFour{{\sc Psi4}}
\newcommand*\Turbomole{{\sc Turbomole}}

%\newref{eq}{name={eq~},names={eqs~},Name={Eq~},Names={Eqs~}}
%\newref{subsec}{name={section~},names={sections~},Name={Section~},Names={Sections~}}

\section{Introduction \label{sec:Introduction}}

Solid-state systems are traditionally modeled with density functional
theory\citep{Hohenberg1964_PR_864,Kohn1965_PR_1133} (DFT) with plane-wave
basis sets of the form $\chi_{\boldsymbol{G}}(\boldsymbol{r})=\Omega^{-1/2}e^{i\boldsymbol{G}\cdot\boldsymbol{r}}$,
where $\boldsymbol{G}$ is a reciprocal lattice vector and $\Omega$
is the volume of the simulation box.\citep{Martin2004__} Importantly,
plane-waves form a systematically improvable basis set, whose accuracy
is determined by a single parameter: the plane-wave kinetic energy
cutoff $E_{\text{cut}}$. The basis set of plane-waves $\boldsymbol{G}$
corresponding to a given cutoff is concisely defined by $\frac{1}{2}\boldsymbol{G}^{2}\le E_{\text{cut}}$,
and the complete basis set limit can in principle be reached by converging
the calculation with respect to $E_{\text{cut}}$. 

However, the plane-wave basis set has a fixed resolution. This is
an issue, since the resolution is the same close to nuclei, where
the electronic wave function undergoes rapid oscillations and where
thereby an extremely fine spatial resolution is needed, as in empty
regions of space where the wave function is typically smooth, lacking
high-frequency components. An accurate description of the core region
requires extremely large values of $E_{\text{cut}}$, resulting in
prohibitive numbers of plane-waves that render calculations untractable.

Plane-wave methods traditionally address this problem by removing
the need to describe the rapid oscillations near the nuclei by employing
various forms of pseudopotentials,\citep{Hamann1979_PRL_1494,Bachelet1982_PRB_4199,Kerker1980_JPCSSP_189,Vanderbilt1985_PRB_8412,Rappe1990_PRB_1227,Troullier1991_PRB_1993,Lin1993_PRB_4174}
a term that we use here in the broadest sense that also includes the
projector-augmented wave (PAW) method.\citep{Bloechl1994_PRB_17953}
These pseudopotentials lead to smooth pseudowave functions, which
can be accurately computed with moderate values of $E_{\text{cut}}$,
thereby enabling powerful applications of DFT to the study of solid-state
systems.\citep{Hasnip2014_PTRSAMPES_20130270}

However, introducing the pseudopotential introduces an approximation,
which may not always be accurate. For instance, it is common practice
to employ pseudopotentials determined for generalized gradient approximation
(GGA) functionals also in calculations using meta-GGA functionals,
even though GGA and meta-GGA functionals do not reproduce the same
core orbitals. The self-consistent use of meta-GGA functionals with
pseudopotentials is an active area of study,\citep{Sun2011_PRB_35117,Yao2017_JCP_224105,Holzwarth2022_PRB_125144,Doumont2022_PRB_195138,Lehtola2023_JCTC_2502}
and fully self-consistent methods for meta-GGA functionals may become
widely available in the future.

Another option for achieving full self-consistency is to avoid the
need for pseudopotentials altogether. For instance, real-space methods
allow employing different levels of resolution in different regions
of space, allowing the use of denser basis sets close to nuclei and
making all-electron calculations tractable.\citep{Lehtola2019_IJQC_25968}
It was also recently pointed out that all-electron calculations could
be made tractable with plane-waves by eliminating the nuclear cusp,
which is hard to describe with plane-waves, by suitable modifications
to the nuclear Coulomb potential.

In \citeref{Gygi2023_JCTC_1300}, \citeauthor{Gygi2023_JCTC_1300}
looked for such a smooth analytic nuclear potential that would be
amenable for all-electron calculations with plane-waves. To guarantee
its accuracy, this potential should yield the exact eigenvalue $E=-1/2$
for the hydrogenic Schrödinger equation
\begin{equation}
-\frac{1}{2r}\frac{{\rm d}^{2}}{{\rm d}r^{2}}r\phi(r)+V(r)\phi(r)=E\phi(r),\label{eq:schrodinger}
\end{equation}
while requiring differentiability of $\phi(r)$ at $r=0$ and the
correct asymptotic limit $\phi(r)\to\exp(-r)/\sqrt{\pi}$ for $r\to\infty$.
\citeauthor{Gygi2023_JCTC_1300}'s solution inverts $V(r)$ from \ref{eq:schrodinger}
using the Ansatz for the $1s$ orbital
\begin{equation}
\phi(r)=\frac{1}{\sqrt{\pi}}e^{-h(r)}\label{eq:ansatz}
\end{equation}
where $h(r)$ is unknown. \citeauthor{Gygi2023_JCTC_1300} finds that
the function
\begin{equation}
h(r;a,b)=r\text{erf}(ar)+b\exp(-a^{2}r^{2})\label{eq:h}
\end{equation}
satisfies the requirements posed above and the arising regularized
potential to be given by 
\begin{equation}
V(r;a,b)=-\frac{1}{2}+\frac{h'(r;a,b)}{r}+\frac{h'(r;a,b)^{2}}{2}+\frac{h''(r;a,b)}{2}.\label{eq:V}
\end{equation}

\Cref{eq:h} has two parameters: $a$ and $b$. \citeauthor{Gygi2023_JCTC_1300}
fixes the $b$ parameter by following \citet{Hamann1979_PRL_1494}
and requiring $\phi(r)$ to be normalized
\begin{equation}
4\pi\int_{0}^{\infty}r^{2}\phi(r)^{2}{\rm d}r=1.\label{eq:normalization}
\end{equation}
Scaling with the nuclear charge $Z$ lead \citeauthor{Gygi2023_JCTC_1300}
to postulate that the potential for $Z>1$ is given by 
\begin{equation}
V(Z;r)=Z^{2}V(Zr).\label{eq:Z-scaling}
\end{equation}

\citeauthor{Gygi2023_JCTC_1300} computed atomic energies for the
H and Be atoms in \citeref{Gygi2023_JCTC_1300} within the local density
approximation (LDA), and found them to be in $\mu E_{h}$ level agreement
with the values of Kotochigova et al.\citep{Kotochigova1997_PRA_191,Kotochigova1997_PRA_5191}
The study then proceeded to plane-wave calculations on various polyatomic
systems---diamond, silicon, MgO, solid argon, and liquid water---where
the convergence of orbital energies, band gaps, ionic forces, and
stress tensors was studied. 

In this contribution, we examine \citeauthor{Gygi2023_JCTC_1300}'s
regularized potential using high-precision atomic calculations including
all electrons. In addition to the LDA, we also consider Hartree--Fock
(HF), generalized gradient approximation (GGA) and meta-GGA level
density functional approximations of total energies. 

The layout of this work is as follows. We begin in \ref{sec:Implementation}
by describing the implementation of the regularized potential in the
\HelFEM{} program,\citep{Lehtola2019_IJQC_25945,Lehtola2019_IJQC_25968,Lehtola2020_PRA_12516,Lehtola2023_JCTC_2502,Lehtola2023_JPCA_4180}
which enables all-electron finite element approaches that routinely
afford sub-$\mu E_{h}$ accuracy in total energies for atoms for a
variety of functionals, and offers a good starting point for studying
the accuracy of \citeauthor{Gygi2023_JCTC_1300}'s regularized potential,
as well. Next, in \ref{sec:totEorbE}, we study the accuracy of total
energies as well as orbital energies and shapes. The computational
details are outlined in \prettyref{subsec:Computational-details},
the accuracy of total energies is studied in \ref{subsec:totE}, and
the examination of the accuracy of orbital energies and shapes is
carried out in \ref{subsec:orbE}. These calculations are carried
out on the He, Be, Ne, Mg, Ar, Ca, Zn, Kr, Sr, Cd, and Xe atoms, which
suffice to study the essential features of the regularized potential.
These results are extended with studies of relative energies in \ref{sec:relE}:
$s$-$d$ excitation energies of first-row transition metal atoms
are studied in \ref{subsec:xcE} and ionization potentials for He--Kr
in \ref{subsec:ionE}. The study concludes in a short summary and
conclusions in \ref{sec:Summary-and-Conclusions}. 

\section{Implementation \label{sec:Implementation}}

We have implemented the potential defined by \cref{eq:h,eq:V,eq:Z-scaling}
in \HelFEM{}. We determine $b(a)$ from \ref{eq:normalization} using
the bisection method and radial quadrature with $N=100$ points with
the default scheme of \citeref{Lehtola2022_JCP_174114}, which is
given by the M3 grid of \citet{Treutler1995_JCP_346} without atomic
size adjustment ($\xi=1$) combined with the Gauss--Chebyshev quadrature
formulas of the second kind of \citet{PerezJorda1994_JCP_6520} that
have simple closed-form expressions, see eqs. (31)--(33) in \citeref{PerezJorda1994_JCP_6520}. 

\citeauthor{Gygi2023_JCTC_1300} tabulated $b(a)$ with 10 decimals
in \citeref{Gygi2023_JCTC_1300}; the values $b(a)$ from our implementation
are compared with \citeauthor{Gygi2023_JCTC_1300}'s in \ref{tab:b-comparison}.
Because of the notable discrepancies observed in the values of $b(a)$---up
to half the decimals disagree---we carried out arbitrary precision
calculations in Maple 2020. We found that employing 20 digit precision
in Maple yielded $b$ converged to 10 decimals. We observe that our
simple scheme yields values for $b$ that are in full agreement with
those from Maple, while tabulation of \citeauthor{Gygi2023_JCTC_1300}---whose
provenance is not described---is not converged to the number of decimals
(10) given in \citeref{Gygi2023_JCTC_1300}, several values only being
accurate to five decimals.

\begin{table*}
\begin{centering}
\begin{tabular}{c|cc|c}
$a$ & $b(a)$, PW & $b(a)$ from \citeref{Gygi2023_JCTC_1300} & $b(a)$, Maple 2020\tabularnewline
\hline 
\hline 
1 & \textbf{3.6442293860}e-01 & \textbf{3.64422938}56e-01 & 3.6442293860e-01\tabularnewline
2 & \textbf{1.9653418941}e-01 & \textbf{1.96534189}82e-01 & 1.9653418941e-01\tabularnewline
3 & \textbf{1.3433604767}e-01 & \textbf{1.34336047}53e-01 & 1.3433604767e-01\tabularnewline
4 & \textbf{1.0200558632}e-01 & \textbf{1.0200558}466e-01 & 1.0200558632e-01\tabularnewline
5 & \textbf{8.2208090847}e-02 & \textbf{8.220809}1118e-02 & 8.2208090847e-02\tabularnewline
6 & \textbf{6.8842562733}e-02 & \textbf{6.88425}55167e-02 & 6.8842562733e-02\tabularnewline
7 & \textbf{5.9213661071}e-02 & \textbf{5.92136}52850e-02 & 5.9213661071e-02\tabularnewline
8 & \textbf{5.1947028410}e-02 & \textbf{5.1947028}250e-02 & 5.1947028410e-02\tabularnewline
9 & \textbf{4.6268541343}e-02 & \textbf{4.62685}59218e-02 & 4.6268541343e-02\tabularnewline
10 & \textbf{4.1708946804}e-02 & \textbf{4.17089}13494e-02 & 4.1708946804e-02\tabularnewline
11 & \textbf{3.7967255428}e-02 & \textbf{3.79672}27308e-02 & 3.7967255428e-02\tabularnewline
12 & \textbf{3.4841536898}e-02 & \textbf{3.48415}73775e-02 & 3.4841536898e-02\tabularnewline
\end{tabular}
\par\end{centering}
\caption{Comparison of $b$ values from the quadrature implementation used
in the present work vs the values given by \citeauthor{Gygi2023_JCTC_1300}
in \citeref{Gygi2023_JCTC_1300}. For comparison, $b$ values solved
with guaranteed precision with Maple 2020 (present work, PW) are also
shown; digits of the two implementations that coincide with the Maple
reference value are shown in bold. \label{tab:b-comparison}}
\end{table*}

\begin{table}
\begin{centering}
\begin{tabular}{c|rr|r}
$a$ & PW & \citeref{Gygi2023_JCTC_1300} & PW, fp\tabularnewline
\hline 
\hline 
1 & 1.676e-13 & 2.118e-11 & 6.075e-17\tabularnewline
2 & 7.479e-13 & -6.321e-11 & 2.601e-16\tabularnewline
3 & -1.786e-13 & 8.788e-12 & 1.175e-16\tabularnewline
4 & 1.368e-13 & 5.335e-11 & 5.450e-17\tabularnewline
5 & -8.680e-15 & -4.950e-12 & -1.042e-16\tabularnewline
6 & 5.135e-15 & 8.574e-11 & 1.495e-16\tabularnewline
7 & -2.430e-15 & 6.176e-11 & 9.851e-17\tabularnewline
8 & -1.722e-15 & 8.355e-13 & 2.504e-16\tabularnewline
9 & -1.731e-15 & -6.773e-11 & 2.306e-16\tabularnewline
10 & 1.136e-15 & 9.429e-11 & 2.547e-16\tabularnewline
11 & -4.674e-16 & 6.101e-11 & 3.285e-16\tabularnewline
12 & 3.256e-16 & -6.268e-11 & 2.372e-16\tabularnewline
\end{tabular}
\par\end{centering}
\caption{Comparison of errors in normalization $\Delta N=4\pi\int_{0}^{\infty}r^{2}\phi(r)^{2}{\rm d}r-1$
of $\phi(r)$ of \prettyref{eq:ansatz} with the values $b$ of \prettyref{tab:b-comparison}
of the present work (PW) and the values of \citeauthor{Gygi2023_JCTC_1300}
in \citeref{Gygi2023_JCTC_1300}, evaluated with Maple 2020 with 25
digits. For comparison, the last column shows the values obtained
using the full precision (fp) value of $b$ with 15 decimals, similarly
to what is used internally in \HelFEM{}. \label{tab:norm-comparison}}
\end{table}

To assess the practical importance of the errors in the $b$ values
used in \citeref{Gygi2023_JCTC_1300}, we have computed the errors
in the normalization arising from the various $b$ values of \ref{tab:b-comparison}
with Maple; these results are shown in \ref{tab:norm-comparison}.
The errors in the normalization of the Ansatz of \ref{eq:ansatz}
are smaller than $10^{-10}$ also with \citeauthor{Gygi2023_JCTC_1300}'s
approximate values for $b$, indicating that the values reported in
\citeref{Gygi2023_JCTC_1300} are likely sufficiently accurate not
to cause severe issues in the validity of the results.

In contrast, if one employs values of $b$ that are really correct
to 10 decimal places, the normalization errors are reduced by a few
orders of magnitude. However, the implementation in \HelFEM{} does
not truncate $b$ to 10 decimal places, but instead determines $b$
to near machine precision. Inserting the value of $b$ printed out
by \HelFEM{} with 15 decimals to Maple shows that $\phi(r)$ is practically
normalized to within machine precision, the largest absolute value
in the rightmost column of \ref{tab:norm-comparison} being 1.5 times
machine epsilon $\epsilon\approx2.2\times10^{-16}$. We therefore
can conclude that our simple scheme to automatically determine $b(a)$
is sufficient to achieve machine precision, and that that pretabulation
of $b(a)$ is thereby not necessary.

\section{Accuracy of Total and Orbital Energies and Shapes \label{sec:totEorbE}}

\subsection{Computational Details \label{subsec:Computational-details}}

Employing the above numerical scheme for finding $b(a)$ in an automated
fashion, we have calculated non-relativistic total energies for HF,
the Perdew--Wang 1992 LDA (PW92),\citep{Bloch1929_ZfuP_545,Dirac1930_MPCPS_376,Perdew1992_PRB_13244}
the Perdew--Burke--Ernzerhof GGA (PBE),\citep{Perdew1996_PRL_3865,Perdew1997_PRL_1396}
as well as the TASKCC meta-GGA functional\citep{Aschebrock2019_PRR_33082,Schmidt2014_JCP_18}
recommended by \citealt{Lebeda2022_PRR_23061} with the normal Coulomb
potential of a point nucleus $E^{\text{point}}$ as well as the regularized
potential of \ref{eq:V} ($E^{\text{regularized}}(a)$) with various
values for the parameter $a$. All density functionals are evaluated
in \HelFEM{} with Libxc.\citep{Lehtola2018_S_1}

We find that the calculations employing the regularized potential
converge more slowly to the complete basis set (CBS) limit than the
calculations with the point nucleus, when the default radial grid
optimized for point nuclei is used. This means that more radial finite
element basis functions are required to reach the CBS limit in calculations
employing the regularized potential.

Following the grid analyses performed in \citerefs{Lehtola2019_IJQC_25945}
and \citenum{Lehtola2023_JPCA_4180}, we considered reoptimizing the
``exponential'' finite element grid\citep{Lehtola2019_IJQC_25945}
\begin{equation}
r_{i}=\left(1+r_{\infty}\right)^{i^{z}/N_{\text{elem}}^{z}}-1,\ i\in[0,N_{\text{elem}}]\label{eq:exploggrid}
\end{equation}
where $r_{\infty}=40a_{0}$ is the employed value for the practical
infinity beyond which all wave functions vanish and $N_{\text{elem}}$
is the used number of elements, by retuning the $z$ parameter that
controls the composition of the grid from the default value $z=2$
optimized for the point nucleus.\citep{Lehtola2019_IJQC_25945,Lehtola2023_JPCA_4180}We
found that calculations with the regularized potential favor denser
grids close to the nucleus than those employing a point nucleus, that
is, large values of $z$ (not shown). We attribute the increased sensitivity
in the region close to the nucleus to the more complicated form of
\ref{eq:V} over the $r^{-1}$ Coulomb interaction. However, grids
with $z\gg2$ tend to lead to poorly convergent self-consistent field
calculations, and we choose to employ the default value $z=2$ also
in the present calculations.

We found that when employing a 10-node Hermite interpolating polynomial
basis,\citep{Lehtola2023_JPCA_4180} which corresponds to employing
a $19^{\text{th}}$ order polynomial scheme, all calculations are
converged to the CBS limit---which we define as 0.1 $\mu E_{h}$
accuracy---when 25 radial elements are employed. For all the systems
studied in this section, calculations with 30 radial elements yield
the same total energy to 7 decimals. We also observe that the regularized
potential results in a lack of a nuclear cusp in the wave function.

\subsection{Accuracy of Total Energies \label{subsec:totE}}

We will next proceed to discuss the errors in total energies caused
by the regularized potential approximation. We define this regularization
error by
\begin{equation}
\Delta E(a)=E^{\text{regularized}}(a)-E^{\text{point}},\label{eq:dE}
\end{equation}
and use it to assess convergence of the total energy to the point
nucleus value. Plots of $\Delta E(a)$ for all studied atoms and functionals
are available in the Supporting Information; we will only present
some of the figures in the main text to exemplify our findings.

\begin{figure}
\begin{centering}
\includegraphics[width=1\linewidth]{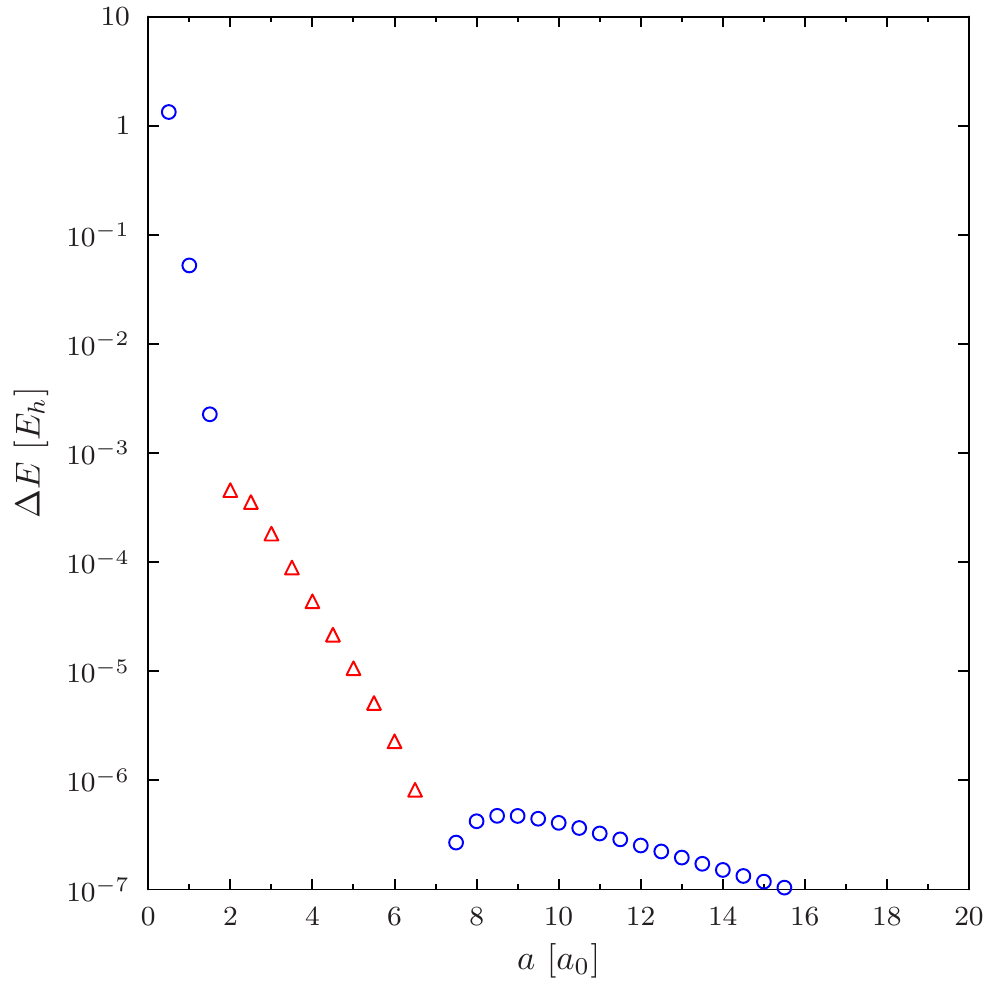}
\par\end{centering}
\caption{Regularization error in the HF total energy of Ne. Note the use of
a logarithmic $y$ axis. Positive energy errors ($E^{\text{regularized}}>E^{\text{Coulomb}})$
shown with blue squares and negative energy errors ($E^{\text{regularized}}<E^{\text{Coulomb}})$
with red triangles.\label{fig:HF-Ne}}
\end{figure}

Depending on the functional, the error in the total energy may be
positive or negative, as is demonstrated by the HF calculation on
Ne in \ref{fig:HF-Ne}. The data in the figure show that there are
sharp minima in $\Delta E(a)$, possibly caused by fortuitous error
cancellation when the structure of the regularized potential matches
the shell structure of the atom. These artefactual error minima may
complicate convergence studies with the regularized potential, but
these issues appear to only affect the lighter atoms. We observe that
total energies can be reproduced to $\mu E_{h}$ accuracy when a sufficiently
large value for $a$ is employed.

\begin{figure}
\begin{centering}
\includegraphics[width=1\linewidth]{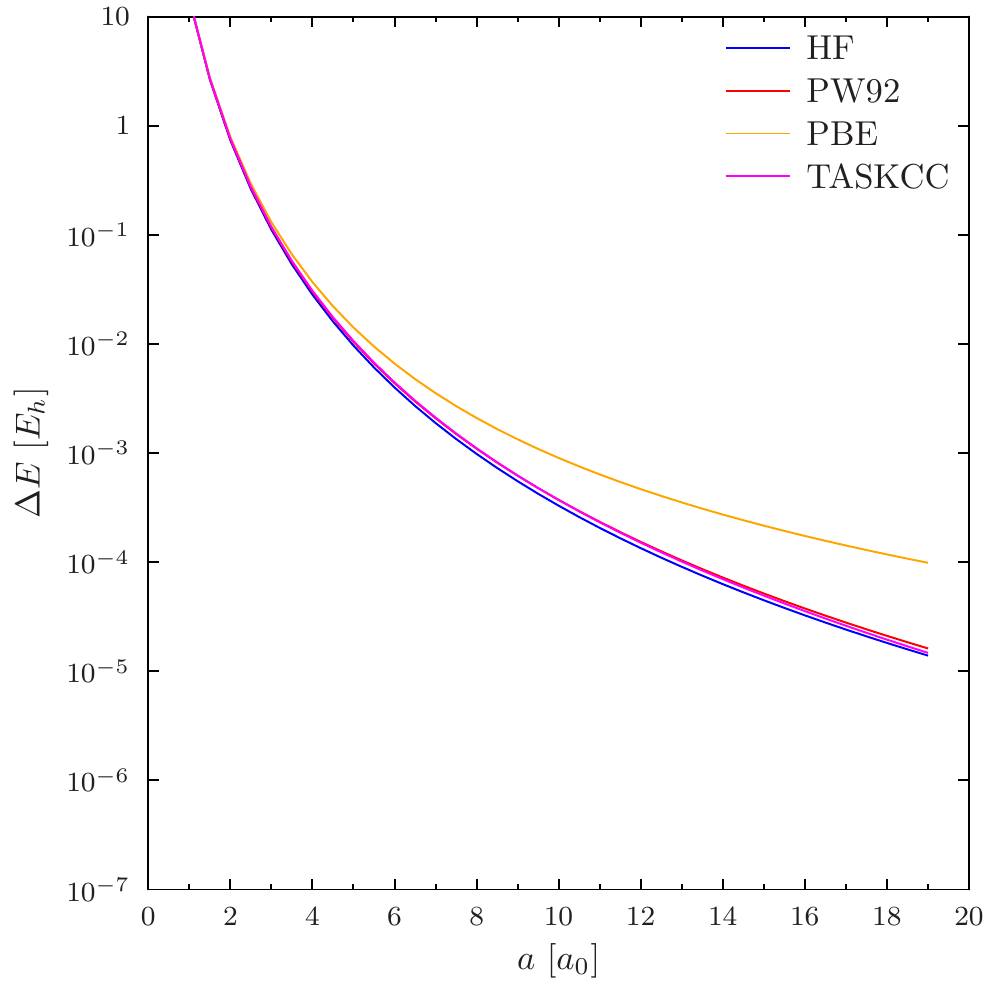}
\par\end{centering}
\caption{Regularization errors in the HF, PW92, PBE, and TASKCC total energies
of Xe. Note the use of a logarithmic $y$ axis. All energy differences
are positive. \label{fig:Xe}}
\end{figure}

Heavier atoms appear to lead to larger differences in total energy.
The differences in total energy $\Delta E(a)$ are positive for all
studied values of $a\in[0.5,19]$ from Mg onwards, and the convergence
plots appear similar for all atoms and functionals. However, the level
of convergence in the total energy depends on the functional. This
is exemplified by the HF, PW92, PBE, and TASKCC calculations on Xe
in \ref{fig:Xe}, which has the typical convex-type form of most of
our results.

We note that the convergence to the CBS limit is slow in $a$. The
data in \ref{fig:Xe} shows that the error decays more slowly with
PBE than with the other studied functionals. The total energy is converged
to 0.1 m$E_{h}$ level accuracy for PBE with the largest value of
$a$ considered in this study ($a=19$), while the differences for
the other functionals are in the tens of microhartrees. For comparison,
\citeref{Gygi2023_JCTC_1300} employed $a=3$ or $a=4$ for non-hydrogen
atoms and up to $a=8$ for hydrogen in polyatomic calculations.

\subsection{Accuracy of Orbital Energies and Shapes \label{subsec:orbE}}

Although total energies converge slowly, we do find that orbital energies
and orbital shapes are indeed accurately captured by the regularized
approximation; tables of orbital energies for all atoms and functionals
are available in the Supporting Information. For example, the errors
in orbital energies of Xe with the TASKCC functional are shown in
\cref{tab:orbE-TASKCC-Xe} for various values of $a$. Even small
values of $a$ that correspond to $E_{h}$ level errors in the total
energy as seen from \ref{fig:Xe} afford accurate orbital energies.
For instance, while $a=2$ reproduces a total energy that differs
by $0.76E_{h}$ from the point nucleus value, the differences in orbital
energies are an order of magnitude smaller. 

\begin{table*}
\begin{tabular}{lrrrrr|r}
Energy  & $ a = 1.0 $ & $ a = 2.0 $ & $ a = 3.0 $ & $ a = 5.0 $ & $ a = 7.0 $ & point nucleus\\
\hline
\hline
$ 1s $& $ -1.3075 $& $ -0.0864 $& $ -0.0126 $& $ -0.0008 $& $ -0.0001 $& $ -1212.0214 $\\
$ 2s $& $ 1.3285 $& $ 0.0370 $& $ 0.0035 $& $ 0.0001 $& $ 0.0000 $& $ -184.6493 $\\
$ 2p $& $ 1.1213 $& $ 0.0662 $& $ 0.0107 $& $ 0.0010 $& $ 0.0002 $& $ -173.6861 $\\
$ 3s $& $ 0.3128 $& $ 0.0090 $& $ 0.0008 $& $ 0.0000 $& $ 0.0000 $& $ -37.9716 $\\
$ 3p $& $ 0.1536 $& $ 0.0118 $& $ 0.0020 $& $ 0.0002 $& $ 0.0000 $& $ -33.3301 $\\
$ 3d $& $ -0.0712 $& $ -0.0046 $& $ -0.0007 $& $ -0.0001 $& $ -0.0000 $& $ -24.6824 $\\
$ 4s $& $ 0.0637 $& $ 0.0017 $& $ 0.0001 $& $ 0.0000 $& $ -0.0000 $& $ -6.9179 $\\
$ 4p $& $ 0.0217 $& $ 0.0019 $& $ 0.0003 $& $ 0.0000 $& $ 0.0000 $& $ -5.2477 $\\
$ 4d $& $ -0.0178 $& $ -0.0011 $& $ -0.0002 $& $ -0.0000 $& $ -0.0000 $& $ -2.3639 $\\
$ 5s $& $ 0.0066 $& $ 0.0001 $& $ -0.0000 $& $ -0.0000 $& $ -0.0000 $& $ -0.7190 $\\
$ 5p $& $ -0.0000 $& $ 0.0000 $& $ 0.0000 $& $ 0.0000 $& $ 0.0000 $& $ -0.3306 $\\
\hline $ \Delta E $ & $ 14.4457645 $ & $ 0.7612456 $ & $ 0.1184265 $ & $ 0.0106325 $ & $ 0.0021111 $ & $ -7233.3416395 $\\
\end{tabular}
\caption{Errors in orbital energies in $ E_h $ for the Xe atom computed with TASKCC and the regularized potential with various values of $ a $. The values obtained with the Coulomb potential of the point nucleus are shown in the last column. For comparison, the last row shows the errors in total energy $ \Delta E $ from the point nucleus value shown in the last column. \label{tab:orbE-TASKCC-Xe} }
\end{table*}

We also find that the shapes of the orbitals are reproduced accurately
already with modest values of $a$. The positions of the orbital density
maxima, defined for radial orbital $\psi_{i}(r)$ as 
\begin{equation}
r_{i}^{\text{max}}=\text{argmax }r^{2}\psi_{i}^{2}(r),\label{eq:densmax}
\end{equation}
are shown in \cref{tab:dmax-TASKCC-Xe} for Xe with the TASKCC functional;
the results for all atoms and functionals are available in the Supporting
Information. Similarly to the orbital energies discussed above, the
positions of the orbital density maxima are already correct to millibohr
with $a=2$. Similar findings also apply to the radial moments of
the orbitals $\langle r^{n}\rangle$ for $n\in[-2,-1,1,2,3]$ (not
shown).

\begin{table*}
\begin{tabular}{lrrrrr|r}
Energy  & $ a = 1.0 $ & $ a = 2.0 $ & $ a = 3.0 $ & $ a = 5.0 $ & $ a = 7.0 $ & point nucleus\\
\hline
\hline
$ 1s $ & $ 0.000115 $ & $ -0.000150 $ & $ -0.000002 $ & $ -0.000000 $ & $ 0.000000 $ & 0.018648\\
$ 2s $ & $ 0.000638 $ & $ 0.000020 $ & $ 0.000002 $ & $ 0.000000 $ & $ 0.000000 $ & 0.102941\\
$ 2p $ & $ 0.000634 $ & $ 0.000037 $ & $ 0.000006 $ & $ 0.000001 $ & $ 0.000000 $ & 0.080418\\
$ 3s $ & $ 0.001579 $ & $ 0.000050 $ & $ 0.000005 $ & $ 0.000000 $ & $ 0.000000 $ & 0.292393\\
$ 3p $ & $ 0.000996 $ & $ 0.000071 $ & $ 0.000012 $ & $ 0.000001 $ & $ 0.000000 $ & 0.278905\\
$ 3d $ & $ -0.000118 $ & $ -0.000011 $ & $ -0.000002 $ & $ -0.000000 $ & $ 0.000000 $ & 0.226757\\
$ 4s $ & $ 0.003284 $ & $ 0.000102 $ & $ 0.000010 $ & $ 0.000000 $ & $ 0.000000 $ & 0.689832\\
$ 4p $ & $ 0.001849 $ & $ 0.000145 $ & $ 0.000025 $ & $ 0.000002 $ & $ 0.000001 $ & 0.706492\\
$ 4d $ & $ -0.000837 $ & $ -0.000059 $ & $ -0.000009 $ & $ -0.000001 $ & $ -0.000000 $ & 0.746263\\
$ 5s $ & $ 0.008732 $ & $ 0.000256 $ & $ 0.000024 $ & $ 0.000001 $ & $ 0.000000 $ & 1.709528\\
$ 5p $ & $ 0.004292 $ & $ 0.000377 $ & $ 0.000067 $ & $ 0.000007 $ & $ 0.000002 $ & 1.937097\\
\end{tabular}
\caption{Errors in positions of orbital density maxima in bohr for the Xe atom computed with TASKCC and the regularized potential with various values of $ a $. The values obtained with the Coulomb potential of the point nucleus are shown in the last column. \label{tab:dmax-TASKCC-Xe} }
\end{table*}

\section{Accuracy of Relative Energies \label{sec:relE}}

\subsection{Accuracy of Excitation Energies \label{subsec:xcE}}

Having established the fast convergence of orbital expectation values
with respect to $a$, one might ask whether the same also holds for
relative energies. In addition to being a stringent check for the
accuracy of density functionals,\citep{Russo1994_JCP_7729,Holthausen2005_JCC_1505,Furche2006_JCP_44103}
the $s$-$d$ excitation energies of first-row transition metals ($s^{2}d^{n-1}\to s^{1}d^{n}$)
are often used to check the reliability of basis sets\citep{Hay1977_JCP_4377,Calaminici2007_JCP_44108}
and pseudopotentials, as they are directly related to the complex
chemistry of transition metals. We note that transition metal systems
were not studied in \citeref{Gygi2023_JCTC_1300}.

Employing spherically symmetric densities for the atoms Sc--Cu in
a spin-unrestricted formulation with 25 radial elements as in \ref{sec:totEorbE},\citep{Lehtola2020_PRA_12516,Lehtola2023_JCTC_2502}
we determine the accuracy of the excitation energies
\begin{equation}
E^{\text{xc}}=E(s^{1}d^{n})-E(s^{2}d^{n-1})\label{eq:exc}
\end{equation}
by computing their difference from the corresponding excitation energies
for a point nucleus
\begin{equation}
\Delta E^{\text{xc}}(a)=E^{\text{xc}}(a)-E^{\text{xc}}(\text{point nucleus}).\label{eq:dExc}
\end{equation}
For reference, approximate point nucleus values are given in \ref{tab:--excitation-energies}.
We find that the $s^{2}d^{n-1}$ and the $s^{1}d^{n}$ states flip
order for small values of $a$ for many atoms, but also that the order
is correctly reproduced when a sufficiently large value of $a$ is
used (not shown).

\begin{table*}
\begin{centering}
\begin{tabular}{c|rrrrrrrrr}
 & Sc & Ti & V & Cr & Mn & Fe & Co & Ni & Cu\tabularnewline
\hline 
\hline 
PW92 & 0.66 & $-0.30$ & $-1.20$ & $-2.05$ & 1.04 & 0.16 & 0.71 & $-1.10$ & 2.40\tabularnewline
PBE & 0.65 & $-0.35$ & $-1.28$ & $-2.17$ & 1.12 & 0.23 & 0.65 & $-1.18$ & 2.38\tabularnewline
TASKCC & 0.84 & $-0.40$ & $-1.56$ & $-2.64$ & 2.03 & 1.07 & 0.17 & 1.39 & 2.59\tabularnewline
r$^{2}$SCAN & 0.52 & $-0.60$ & $-1.66$ & $-2.67$ & 1.87 & 0.76 & 0.38 & $-1.27$ & 2.59\tabularnewline
\end{tabular}
\par\end{centering}
\caption{$s$-$d$ excitation energies in eV for point nuclei from spin-unrestricted
calculations employing spherical densities. \label{tab:--excitation-energies}}
\end{table*}

\begin{figure}
\begin{centering}
\includegraphics[width=1\linewidth]{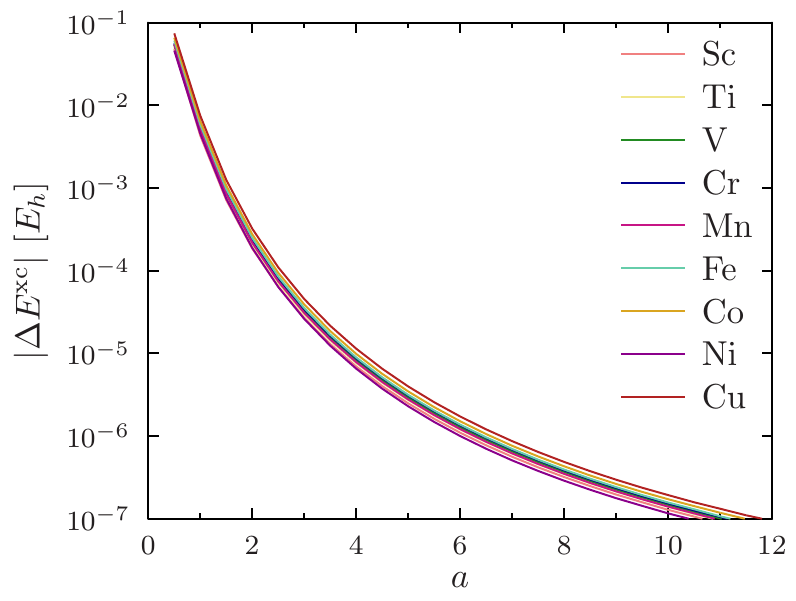}
\par\end{centering}
\caption{Convergence of $s$-$d$ excitation energies $E^{\text{xc}}$ with
decreasing regularization parameter $a$. Results shown for the PBE
functional, other functionals yield analogous results. \label{fig:Convergence-of--}}

\end{figure}

We observe that $\Delta E^{\text{xc}}$ often has the same sign for
the studied range of $a$, implying monotonic convergence of the excitation
energy, but also that some exceptions also exist where the sign of
the error changes at a small value of $a$ (not shown). For this reason,
it suffices to demonstrate the rapid convergence of $|\Delta E^{\text{xc}}|$,
shown in \ref{fig:Convergence-of--} for the PBE functional, as PW92,
TASKCC and r$^{2}$SCAN were found to yield similar results (not shown). 

As expected, the data for all atoms Sc--Cu appear similar. At small
$a$, the potential for erroneous state orderings is proved by the
error in the excitation energy shown in \ref{fig:Convergence-of--}
being in the order of eV, that is, of the same order of magnitude
as the point-nucleus excitation energies themselves (\ref{tab:--excitation-energies}).
However, one can also observe from \ref{fig:Convergence-of--} that
already the value $a=4$ appears to afford errors in the order of
$\mathcal{O}(10^{-5}E_{h})$, that is, sub-meV level precision for
excitation energies.

\subsection{Accuracy of Ionization Potentials \label{subsec:ionE}}

\begin{figure}
\begin{centering}
\includegraphics[width=0.95\linewidth]{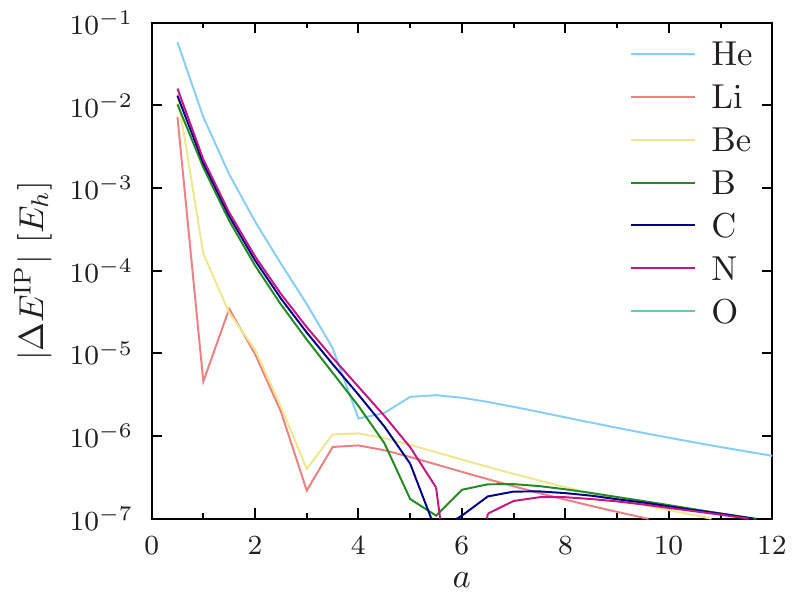}
\par\end{centering}
\caption{Regularization errors in the PBE ionization potential for He--O as
a function of the regularization parameter $a$. \label{fig:He-O}}
\end{figure}

\begin{figure}
\begin{centering}
\includegraphics[width=0.95\linewidth]{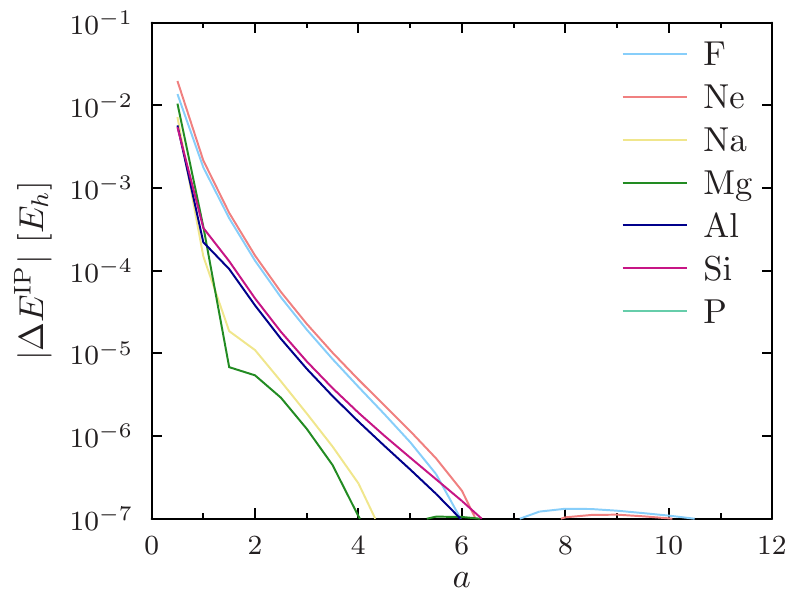}
\par\end{centering}
\caption{Regularization errors in the PBE ionization potential for F--P as
a function of the regularization parameter $a$. \label{fig:F-P}}
\end{figure}

\begin{figure}
\begin{centering}
\includegraphics[width=0.95\linewidth]{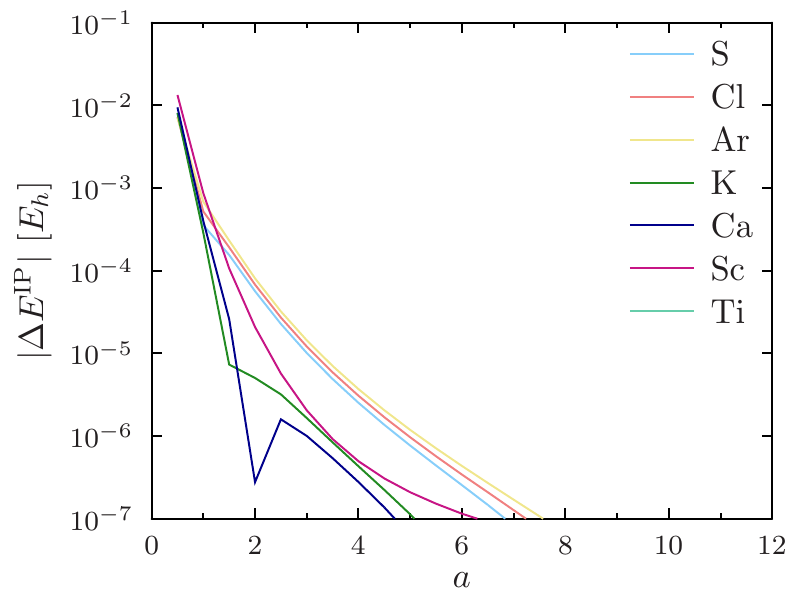}
\par\end{centering}
\caption{Regularization errors in the PBE ionization potential for S--Ti as
a function of the regularization parameter $a$. \label{fig:S-Ti}}
\end{figure}

\begin{figure}
\begin{centering}
\includegraphics[width=0.95\linewidth]{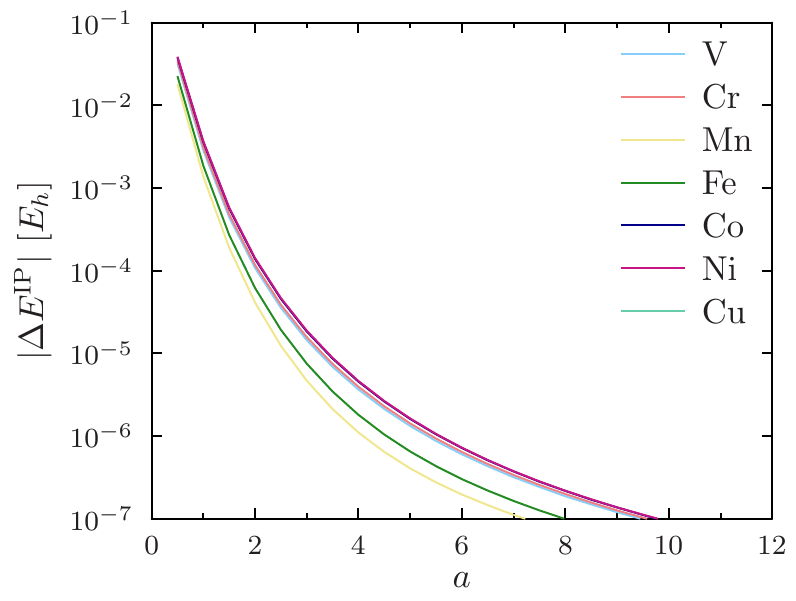}
\par\end{centering}
\caption{Regularization errors in the PBE ionization potential for V--Cu as
a function of the regularization parameter $a$. \label{fig:V-Cu}}
\end{figure}

\begin{figure}
\begin{centering}
\includegraphics[width=0.95\linewidth]{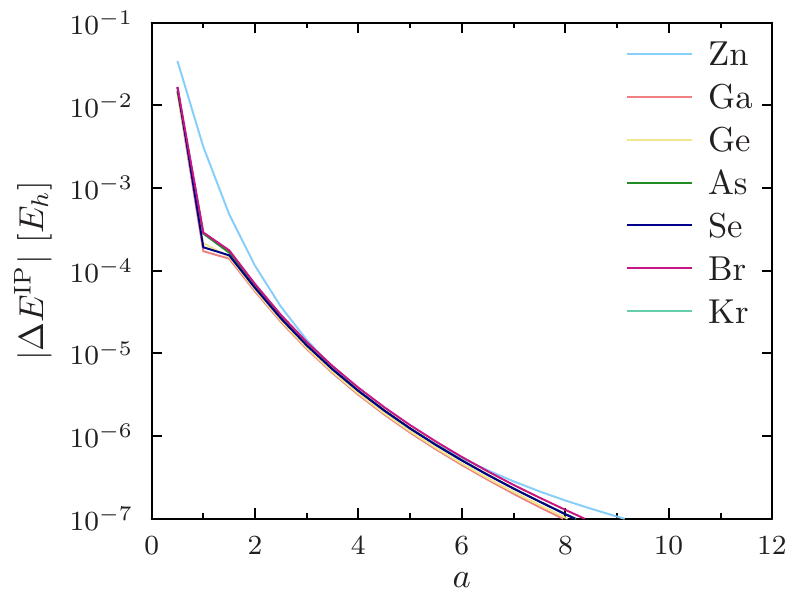}
\par\end{centering}
\caption{Regularization errors in the PBE ionization potential for Zn--Kr
as a function of the regularization parameter $a$. \label{fig:Zn-Kr}}
\end{figure}

Having established the accuracy of $s$-$d$ excitation energies,
we continue by examining the accuracy of ionization potentials for
He--Kr. Also these calculations employ 25 radial elements. Analogously
to \ref{subsec:xcE}, we employ a spin-unrestricted formalism with
spherically symmetric densities to compute the ionization potential
\begin{equation}
E^{\text{IP}}=E(\text{cation})-E(\text{neutral}).\label{eq:Eip}
\end{equation}
The errors in the ionization potential 
\begin{equation}
\Delta E^{\text{IP}}(a)=E^{\text{IP}}(a)-E^{\text{IP}}(\text{point nucleus})\label{eq:dEip}
\end{equation}
 are shown for the PBE functional in \ref{fig:He-O} for He--O, in
\ref{fig:F-P} for F--P, in \ref{fig:S-Ti} for S--Ti, in \ref{fig:V-Cu}
for V--Cu, and in \ref{fig:Zn-Kr} for Zn--Kr. The other studied
functionals (PW92, TASKCC and r$^{2}$SCAN) again yielded similar
results (not shown).

The ionization potential for the helium atom converges surprisingly
slowly with increasing $a$. However, this is easily understood, as
the ionization potential of He is really a core property: it depends
explicitly on the 1s orbital. The ionization potentials of heavier
atoms converge more rapidly to sub-$\mu E_{h}$ precision. 

One can again observe in \crefrange{fig:He-O}{fig:Zn-Kr} that $a=4$
affords sub-meV precision of $\mathcal{O}(10^{-5}E_{h})$ for ionization
potentials in all cases, including He.


\section{Summary and Conclusions \label{sec:Summary-and-Conclusions}}

We have thoroughly examined the regularized nuclear potential recently
suggested by \citet{Gygi2023_JCTC_1300}. We have discussed the determination
of the $b$ parameter in the potential based on the strength $a$
of the regularization, and described a simple method to determine
values of $b(a)$ that satisfy the normalization condition to machine
precision. We implemented the potential in the \HelFEM{} program,\citep{Lehtola2019_IJQC_25945,Lehtola2019_IJQC_25968,Lehtola2020_PRA_12516,Lehtola2023_JCTC_2502,Lehtola2023_JPCA_4180}
which we used to carry out a series of atomic calculations to sub-$\mu E_{h}$
precision with the PW92, PBE, TASKCC, and r$^{2}$SCAN functionals.

We studied the convergence of total energies, orbital energies, and
orbital shapes of closed-shell atoms from Ne to Xe, as well as $s$-$d$
excitation energies of Sc--Cu and the ionization potentials of He--Kr.
We found that although the total energies converge slowly with $a$,
exhibiting differences from the point nucleus value of the order of
0.1 m$E_{h}$ with $a=19$, orbital energies and shapes converge much
more rapidly, exhibiting small errors already with $a=5$. The $s$-$d$
excitation energies and ionization potentials likewise showed much
faster convergence to the point nucleus limit with increasing $a$
than the total energies, reaching sub-meV precision with $a=4$. 

These results lend independent support to the accuracy of \citeauthor{Gygi2023_JCTC_1300}'s
regularized potential. Although the regularized potential can result
in non-monotonic convergence with respect to $a$, as demonstrated
by total energies that can either overestimate or underestimate the
point-nucleus value, the rapidity in which many observables converge
to the point nucleus values suggest that the regularized potential
indeed appears to offer a tractable and reliable way to approach all-electron
calculations with plane-waves.

\section*{Supporting Information}

Convergence plots and tables of orbital energies and orbital electron
density maxima for all studied atoms and all studied functionals.

\section*{Acknowledgments}

We thank the Academy of Finland for financial support under project
numbers 350282 and 353749. We thank CSC -- IT Centre for Science
(Espoo, Finland) for computational resources.

\clearpage{}

\bibliography{citations}

\end{document}